\documentclass[letterpaper]{article}

\usepackage{natbib, alifeconf}  
\usepackage{url}

\usepackage{cleveref}
\title{Glaberish: Generalizing the Continuously-Valued Lenia Framework to Arbitrary Life-Like Cellular Automata}

\author{Q. Tyrell Davis and Josh Bongard \thanks{$^\ast$ University of Vermont, [qdavis, jbongard]@uvm.edu} }

\begin{document}
\maketitle

\begin{abstract}

 Recent work with Lenia, a continuously-valued cellular automata (CA) framework, has yielded $\sim$100s of compelling, bioreminiscent and mobile patterns. Lenia can be viewed as a continuously-valued generalization of the Game of Life, a seminal cellular automaton developed by John Conway that exhibits complex and universal behavior based on simple birth and survival rules. Life's framework of totalistic CA based on the Moore neighborhood includes many other interesting, Life-like, CA. A simplification introduced in Lenia limits the types of Life-like CA that are expressible in Lenia to a specific subset. This work recovers the ability to easily implement any Life-like CA by splitting Lenia's growth function into genesis and persistence functions, analogous to Life's birth and survival rules. We demonstrate the capabilities of this new CA variant by implementing a puffer pattern from Life-like CA Morley/Move, and examine differences between related CA in Lenia and Glaberish frameworks: Hydrogeminium natans and s613, respectively. These CA exhibit marked differences in dynamics and character based on spatial entropy over time, and both support several persistent mobile patterns. The CA s613, implemented in the Glaberish framework, is more dynamic than the Hydrogeminium CA (and likely most Lenia-based CA) in terms of a consistently high variance in spatial entropy over time. These results suggest there may be a wide variety of interesting CA that can be implemented in the Glaberish variant of the Lenia framework, analogous to the many interesting Life-like CA outside of Conway's Life. Supporting information and resources are open-source\footnote{Links to supporting resources, including notebooks for replicating this paper's figures and additional animations are consolidated at \url{https://rivesunder.github.io/yuca}}.

\end{abstract}

\begin{figure}[ht]                                                               
\begin{center}                                                              
  \includegraphics{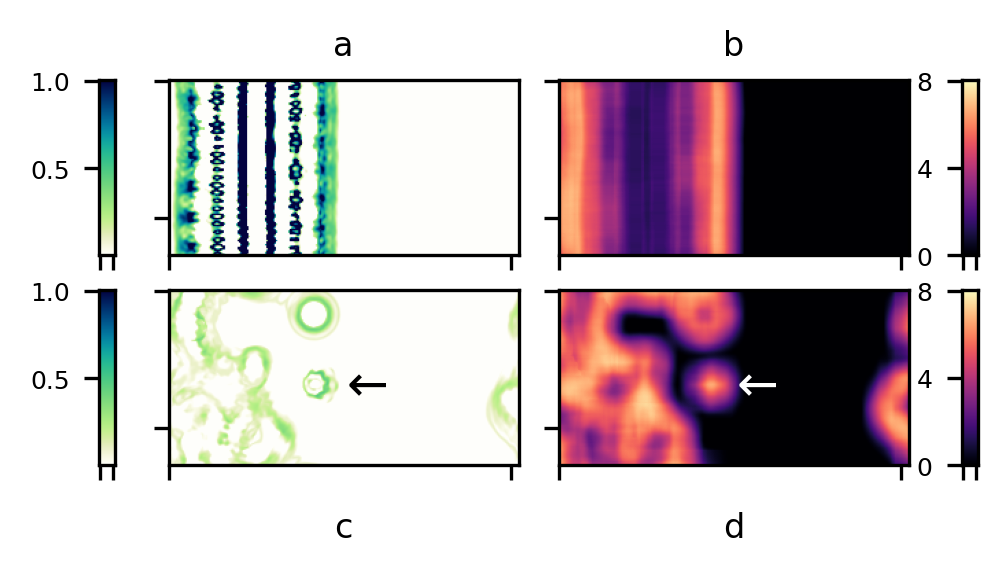}  
  \caption{Hydrogeminium natans (Lenia) and s613 (Glaberish) CA after 47 time steps. Both CA started with the same 64-wide vertical strip of cells with random uniform intial values. a) Hydrogeminium grid state after 47 time steps. b) Spatial entropy map of a. c) s613 grid state after 47 time steps, with emerged ``frog" glider pattern. d) Spatial entropy map of c. Arrows call out glider pattern in s613. Spatial entropy was computed with a window size of 23, entropy units are in bits.}
  \label{fig:teaser}                                                   
\end{center}                                                                
\end{figure}

\section{Introduction}

John Conway's Game of Life is a cellular automaton (CA, plural: cellular automata) and archetypical example of a complex and computationally universal system arising from simple rules \citep{gardner1970, berlekamp2004}. Life is a ``zero-player" (or simulation) game invented by Conway during coffee breaks with help from colleagues; its invention laid the foundations for a vibrant community of hobbyists and professional researchers to study Life and related CA, with some of the most interesting and beautiful work in CA the work of non-academic artists, engineers, and tinkerers. Life consists of binary cells on a grid, with each cell's dynamics defined by the sum of neighbors and the cell's own state. The dynamics of Life depend entirely on cell states and neighborhood values and follow a simple set of rules. 

Life is one of 262,144 possible rules in the framework of Life-like CA. Rules can be written as Bx/Sy, where x and y are any, none, or all integers from 0 through 8. Life is written as B3/S23: empty cells with 3 neighbors are {\bfseries B}orn, or become 1, and active cells with 2 or 3 neighbors {\bfseries S}urvive, and retain a state of 1. Life-like CA update synchronously and discretely. Because B rules and S rules are 9-bits each, and can be implemented in any combination, there are $2^9 \cdot 2^9 = 2^{18}$ possible rule combinations of Life-like CA. 

In Life cells can have a state of 0 or 1, and each cell has a neighborhood consisting of the cells orthogonally and diagonally adjacent to it. With exactly 3 neighbors cell state becomes 1. With 2 neighbors, cell state remains unchanged, regardless of whether the current state is 0 or 1.
All other cells take on a state of 0. Note that it is possible to write a description of Life rules without referring to the particular value of the cell state. This distinction enables facile implementation of Life in the Lenia framework, but is not a general characteristic for every Life-like CA. 


  


Inspired by Life, CA frameworks have subsequently been expanded to larger neighborhoods \citep{Evans2001, pivato2007}, higher dimensions \citep{bays1987, chan2020}, multiple channels \citep{chan2020}\footnote{See also \url{https://softologyblog.wordpress.com/2018/03/09} and \url{https://github.com/slackermanz/vulkanautomata} }, and continuous-value states and updates \citep{rucker2003, chan2019}\footnote{See also \url{https://github.com/rudyrucker/capow},
\url{https://www.rudyrucker.com/capow}, and \url{https://arxiv.org/abs/1111.1567}} to give just a few examples of the many Life-inspired CA projects.

An early framework for continuous CA, developed by Rudy Rucker in the 1990s, was CAPOW \citep{rucker2003}. Later, Stephan Rafler described the SmoothLife CA based on sharply defined inner and outer neighborhoods, with birth and survival rules defined by a pair of intervals comprised of smooth step functions\footnote{Rafler, S. (2011). Generalization of Conway’s ``Game of Life" to a continuous domain - SmoothLife. pre-print on ArXiv \url{https://arxiv.org/abs/1111.1567}}. Bert Chan developed the continuous Lenia CA framework with a different formulation: neighborhoods are defined by smooth convolution kernels and a single update function (called the growth function) \citep{chan2019}.

Neither SmoothLife nor Lenia explicitly considered Life-like CA in general: SmoothLife was developed with a single continuous interval each for birth and survival ({\it i.e.} SmoothLife used ``Bays space" rules \citep{bays1987}), and Lenia updates do not depend on cell state. This potentially leaves a vast volume of continuous CA with interesting dynamics that are not readily implementable in Lenia or SmoothLife. 

A Life-like CA of particular interest is called Move or Morley, with rules B368/S245 (Figure \ref{fig:morley_in_lenia}). Morley supports a commonly occurring puffer, a mobile pattern that continuously generates persistent patterns along its trajectory. The Morley puffer is a simple example of an infinite growth pattern, of particular interest as an indicator of complexity in CA. Morley has multiple B and S conditions, none of which overlap and most of which are not contiguous, making it an ideal Life-like CA for exploring the limitations of previous continuous CA frameworks. 

In the next section we demonstrate Lenia does not have a simple implementation of Life-like CA Morley. This limitation is exemplified in Figures \ref{fig:morley_in_lenia} and \ref{fig:timelapse}. Choosing to view this as a problem, we  incorporate conditional updates based on cell state, and name the resulting CA framework {\it Glaberish}, a nod to the Latin etymology of Lenia and a reference to the role of Life-like CA beyond Conway's Life in studying complex systems. Glaberish takes its root from another Latin word for smooth\footnote{Lenia and Glaberish are based on the Latin root words {\it lenis} and {\it glaber}, respectively.}, and Glaberish is to Lenia as Life-like CA are to Conway's Life.

In Results several techniques from the literature are used to assess CA implemented in the Lenia framework and in Glaberish. A spatial entropy metric, similar to that in \citep{wuensche1999}, is shown applied to CA grid states in Lenia and Glaberish in Figure \ref{fig:teaser}, with an emergent glider pattern in the Glaberish CA (s613). This metric is further considered in the context of previous work in the Discussion section. Speculation about the potential value of increasingly complicated systems, {\it i.e.} continuous CA, forms the basis for possible future research directions discussed in our Conclusions.

\section{Glaberish is Lenia with conditional updates } \label{sec:workup}

\begin{figure}[t]                                                              
\begin{center}                                                              
  \includegraphics{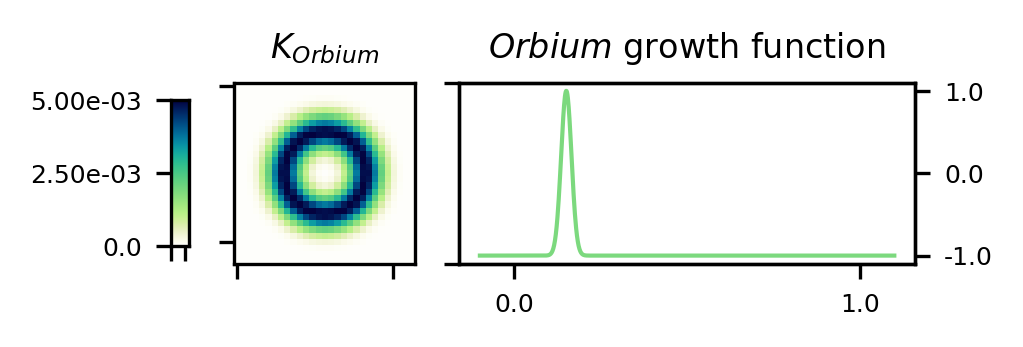}    
  \caption{Lenia neighborhood kernel and growth function for Orbium CA. A neighborhood kernel and a growth function gives rise to complex behavior in Lenia. The Gaussian kernel defined by $(\mu=0.5, \sigma=0.15)$ and Gaussian growth function defined by $(\mu=0.15, \sigma=0.015)$, shown here, supports the exemplary {\it Orbium} glider pattern.}
  \label{fig:orbium_rules}
  \label{fig:standard_lenia}
\end{center}        
\end{figure}  

Like Life, Lenia defines CA dynamics with a neighborhood and an update function. Neighborhood values are the result of 2D convolution with kernel $K$, which become input to growth function $G$. CA dynamics in the Lenia system are defined as

\begin{equation}
    A^{t+dt} = \psi ( A^{t)}+ dt \cdot G(K \ast A^t) )     
    \label{eqn:standard_lenia}
\end{equation}          

where $\psi$ is a squashing or clipping function that limits values to between 0.0 and 1.0, and $A^{t}$ is the CA grid at time $t$. 

The kernel $K$ and growth function $G$ can take various forms, but a typical Lenia CA uses Gaussian functions of the form
       
\begin{equation}
    f(x) =  exp\left( - \left(\frac{(x-\mu)}{2\sigma}\right)^2 \right)
    \label{eqn:gaussian}
\end{equation}              

Figure \ref{fig:standard_lenia} visualizes $K$ and the growth function for a Lenia CA, Orbium. The neighborhood kernel is a $(\mu=0.5, \sigma=0.15)$ Gaussian with a 2D grid of radial distance from center as input $x$. $K$ is normalized to have a sum of 1.0, and the output of convolution with $K$ is input for the growth function. In the same CA, the growth function $G(\cdot)$ is a Gaussian  with $(\mu=0.15, \sigma=0.015)$. Note that for use as a growth function, $f(x)$ is stretched to yield values from -1 to 1, {\it i.e.} $G(x) = 2 \cdot f(x) - 1$. The Orbium CA supports an iconic glider pattern of the same name, described in \citep{chan2019}. 

\begin{figure}[t]                                           
\begin{center}
  \includegraphics{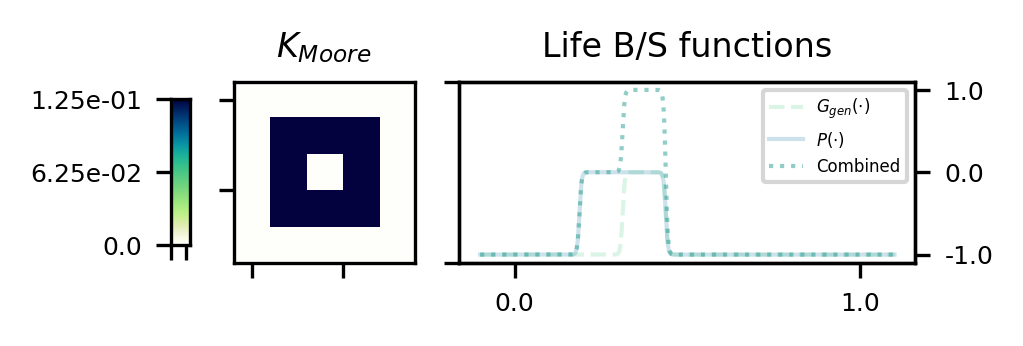}  
  \caption{Conway's Game of Life implemented in the Lenia framework. Kernel $K$ is a Moore neighborhood, and a step-wise growth function corresponds to B3/S23.}
  \label{fig:life_in_lenia}                           
\end{center}                                          
\end{figure}

\begin{figure}[t]                                           
\begin{center}
  \includegraphics{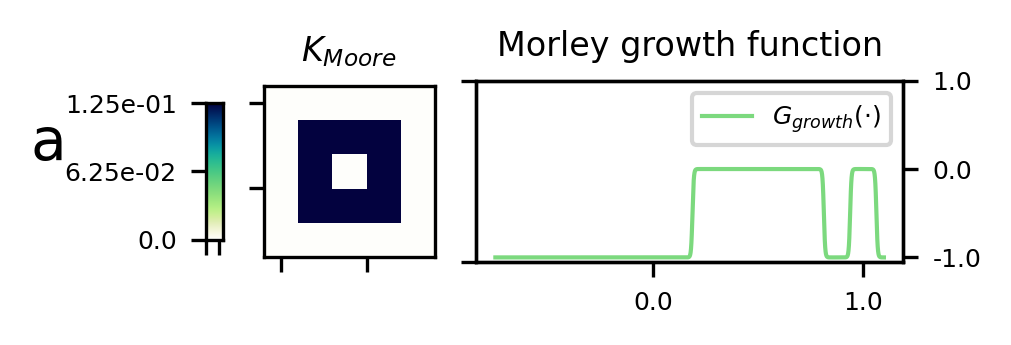}  
  \includegraphics{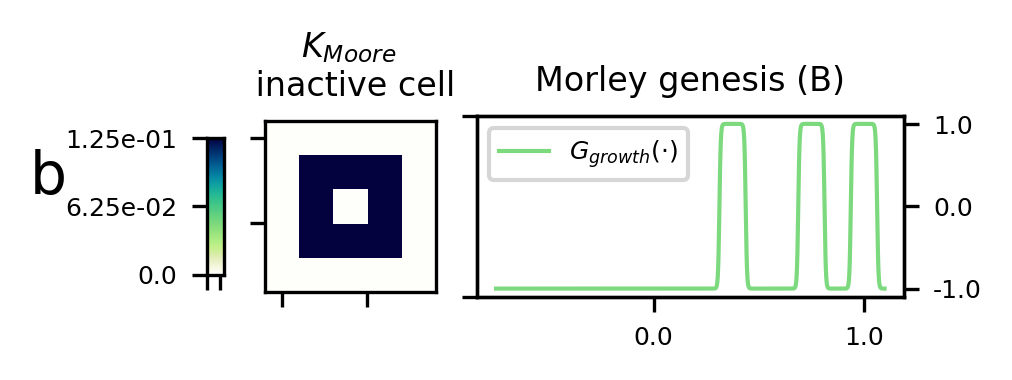}  
  \includegraphics{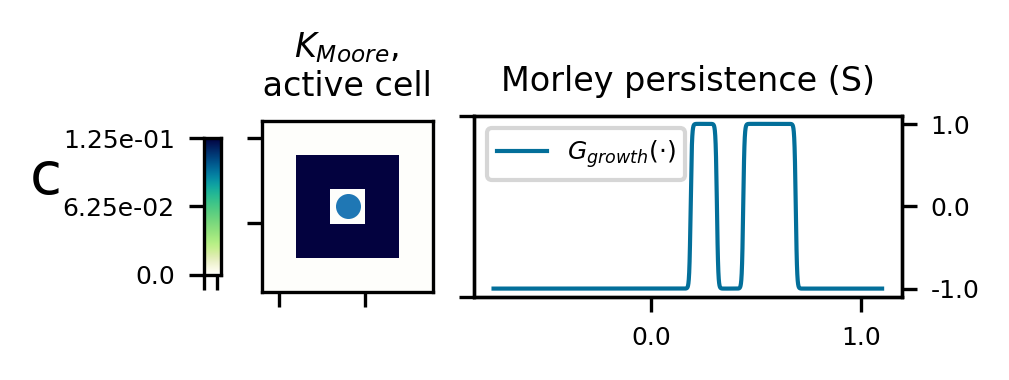}  
  \caption{Continuous implementations of the Life-like Morley/Move CA (B368/S245). a) Implemented in Lenia with updates defined by a single growth function. b) and c) Implemented in Glaberish with separate genesis and persistence functions.}
  \label{fig:morley_in_lenia}                           
\end{center}                                          
\end{figure}


Figure \ref{fig:life_in_lenia}a shows how Conway's Life can be implemented in the Lenia framework. The B and S rules overlap, so if we sum the corresponding continuous intervals we get a two-step staircase function; the default value is -1, cells go to 0, values in the survival interval yield a value of 0, no change, and values in the birth interval yield an update of +1. Growth occurs only where B and S rules overlap. 

Lenia can only readily implement Life-like CA where values in the B interval are a subset of those in the S interval, to enable growth and avoid survival outside of survival intervals. B368/S245, aka Morley (Figure \ref{fig:morley_in_lenia}), is an example of a CA with no overlap between B and S rules and is not readily implementable in the Lenia framework. Combining the B/S rules as in Figure \ref{fig:life_in_lenia} results in a growth function that never returns values greater than 0 (Figure \ref{fig:morley_in_lenia}a), yielding a CA that is not capable of growth. 

To extend Lenia to a full generalization of Life-like CA, we add conditional updates, splitting the growth function into genesis and persistence functions that depend on the cell state. Equation \ref{eqn:glaberish} defines the Glaberish update.

\begin{equation}
    A^{t + dt} = \psi \left( \\
    A^t +  dt [ (1 - A^t)G_{gen}(N) + A^t  P(N) ] 
    \\ \right)
    \label{eqn:glaberish}
\end{equation} 
                
\begin{figure}[ht]                                                           
\begin{center}                                                                  
  \includegraphics{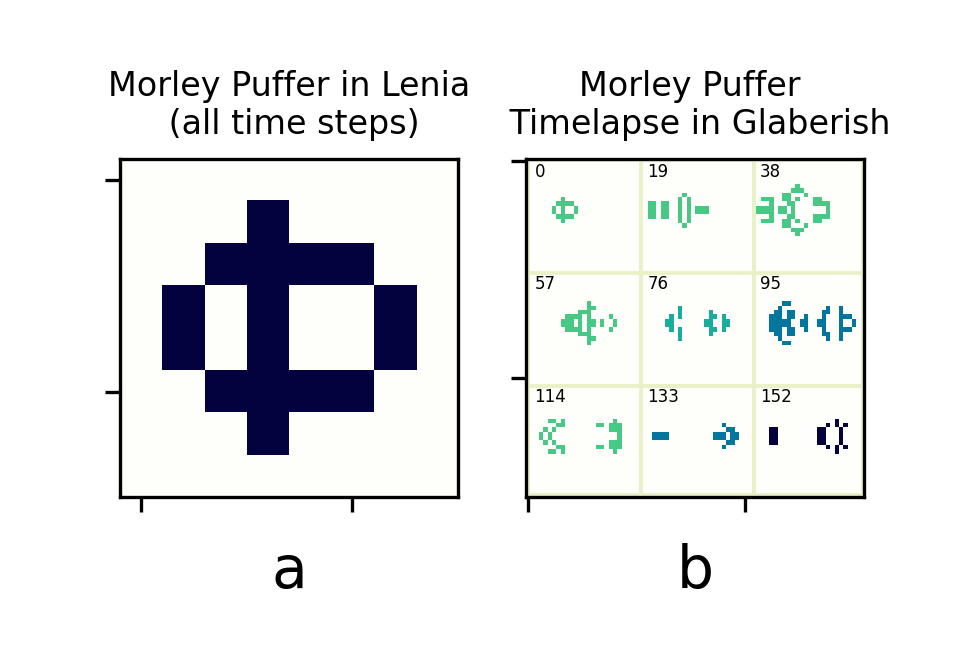}          
  \caption{A common puffer in Life-like CA Morley/Move (B368/S245). a) The puffer pattern is static in Lenia, under the growth function in Figure \ref{fig:morley_in_lenia}a. b) Implemented in Glaberish, the pattern exhibits expected behavior: moving across the grid while leaving a trail of oscillating patterns.}
  \label{fig:timelapse}                                                         
\end{center}                                                                    
\end{figure}                                                                    

In Equation \ref{eqn:glaberish} we use $N$ to denote the result of neighborhood convolution $K \ast A^t$. The growth function $G(\cdot)$ is replaced in Equation \ref{eqn:glaberish} by genesis function $G_{gen}(\cdot)$ and persistence function $P(\cdot)$.
Conditional update dynamics rescue the ability to implement B368/S245, and a timelapse of the common Morley puffer pattern is shown in Figure \ref{fig:timelapse}.

\section{Results} 

\label{sec:results}


\begin{figure}[ht]                                                               
\begin{center}                                                              
  \includegraphics{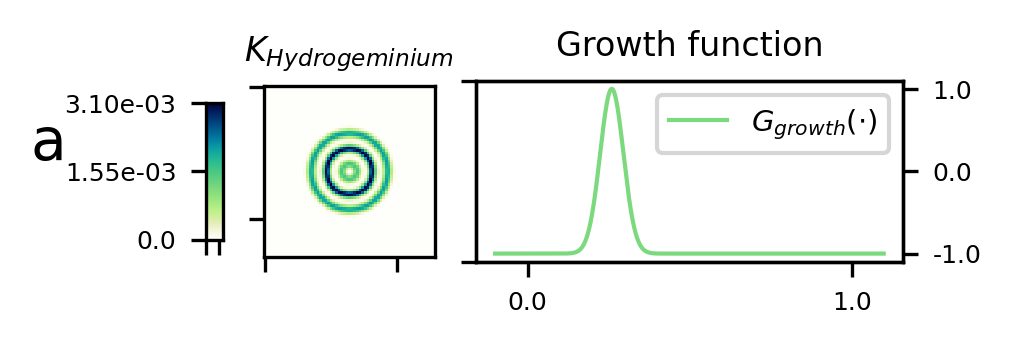}  
  \includegraphics{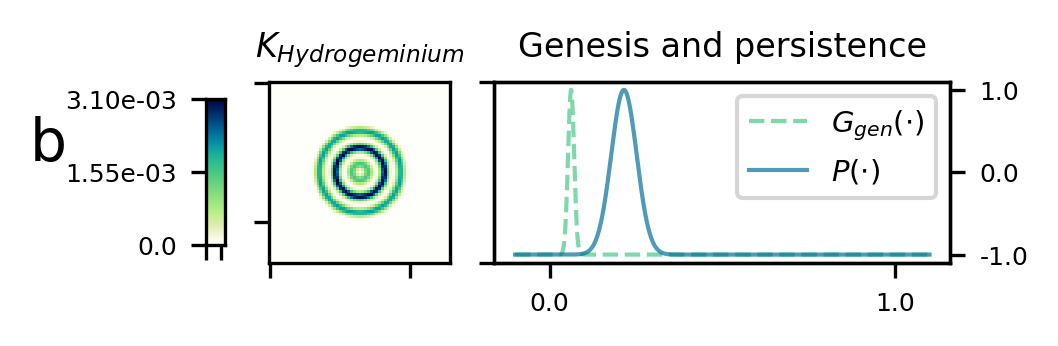}  
  \caption{Rule visualization in Lenia and Glaberish. a) Hydrogeminium, a CA based on the Lenia framework. b) s613, a Glaberish CA. These use the same neighborhood kernel but differ in their update functions. Most notably s613 splits updates into genesis and persistence dependent on current cell state. Despite minor differences in the update functions and marked differences in CA dynamics, both systems readily support mobile patterns. }
  \label{fig:gem_s613_updates}                                                   
\end{center}                                                                
\end{figure}

We compare the Hydrogeminium CA from Lenia  \citep{chan2019} to an evolved CA rule set under the Glaberish framework. We refer to the Glaberish CA as ``s613" for the random seed used to evolve it, selecting for poor performance of neural networks trained to predict whether all cell states fall to 0 after a set number of time steps, described in \citep{davis2022}. These CA share a neighborhood kernel, and the persistence function from s613 resembles a slightly shifted version of the growth function from Hydrogeminium natans (Figure \ref{fig:morley_in_lenia}). The most notable difference between the two sets of rules, the s613 genesis function (a Gaussian with $\mu = 0.063$ and $\sigma = 0.0088$) makes for markedly different dynamics between the two CA.

Hydrogeminium is defined by a (shared with s613) neighborhood kernel with three rings, the weighted sum of three Gaussians acting on the radial distance from center in a 2D grid, with parameters $(\mu, \sigma)$ = $[(0.0938, 0.033), (0.2814, 0.0330), (0.469, 0.033)]$ and weights $[0.5, 1.0, 0.667]$. The growth function has parameters $(\mu, \sigma)$ = $(0.26, 0.036)$. s613 shares the neighborhood function (as well as the same step size $dt=0.1$), but has genesis and persistence functions with parameters $(\mu, \sigma) = (0.0621, 0.0088)$ and $(0.2151, 0.0369)$, respectively.

\begin{figure}[ht]                                                       
\begin{center}                                                    
  \includegraphics{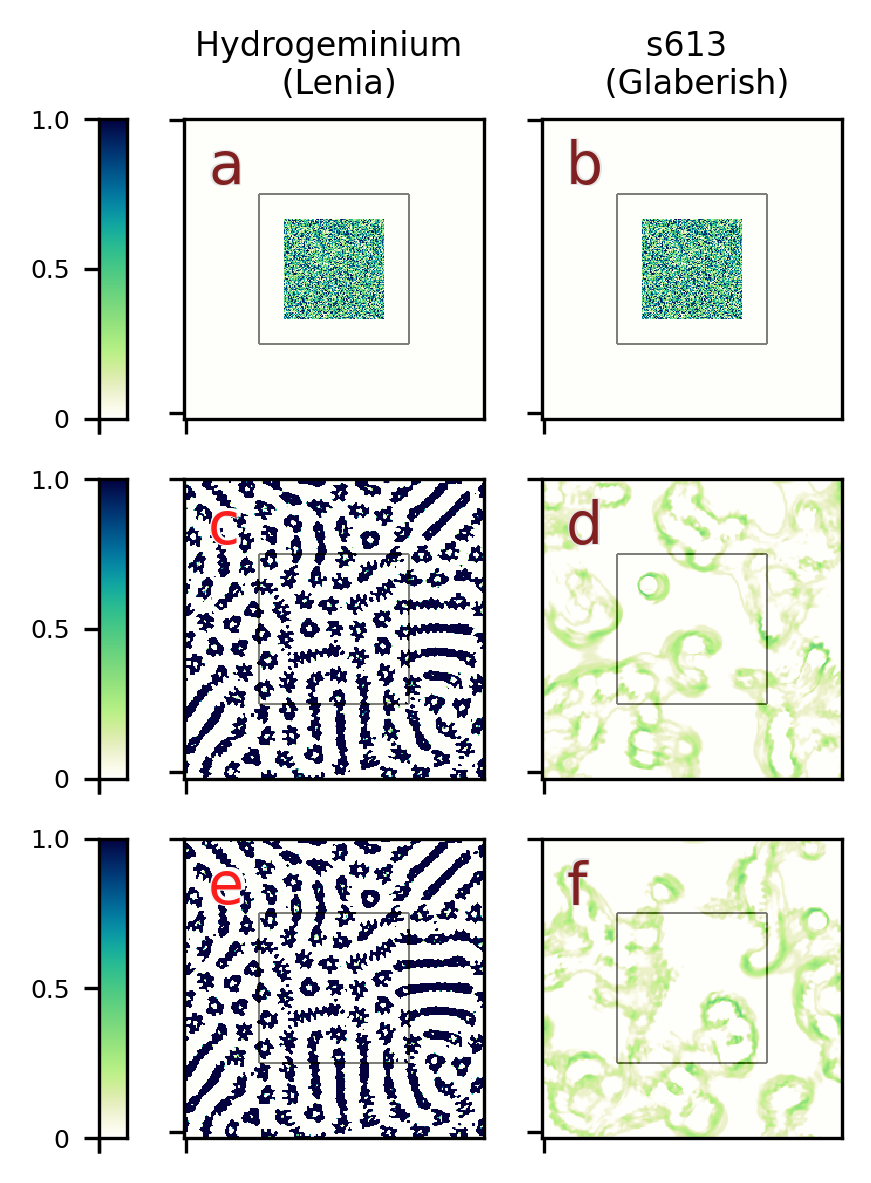}
  \caption{Grid state progression from random uniform initial states. a) Hydrogeminium at step 0. c) Hydrogeminium at step 512. c) Hydrogeminium at step 1024. b) s613 at step 0 (same as in \ref{fig:gem_s613_compare}a). d) s613 at step 512. f) s613 at step 1024.}
  \label{fig:gem_s613_compare}                                      
\end{center}                                                            
\end{figure}

\begin{figure}[ht]                                                       
\begin{center}                                                            
  \includegraphics{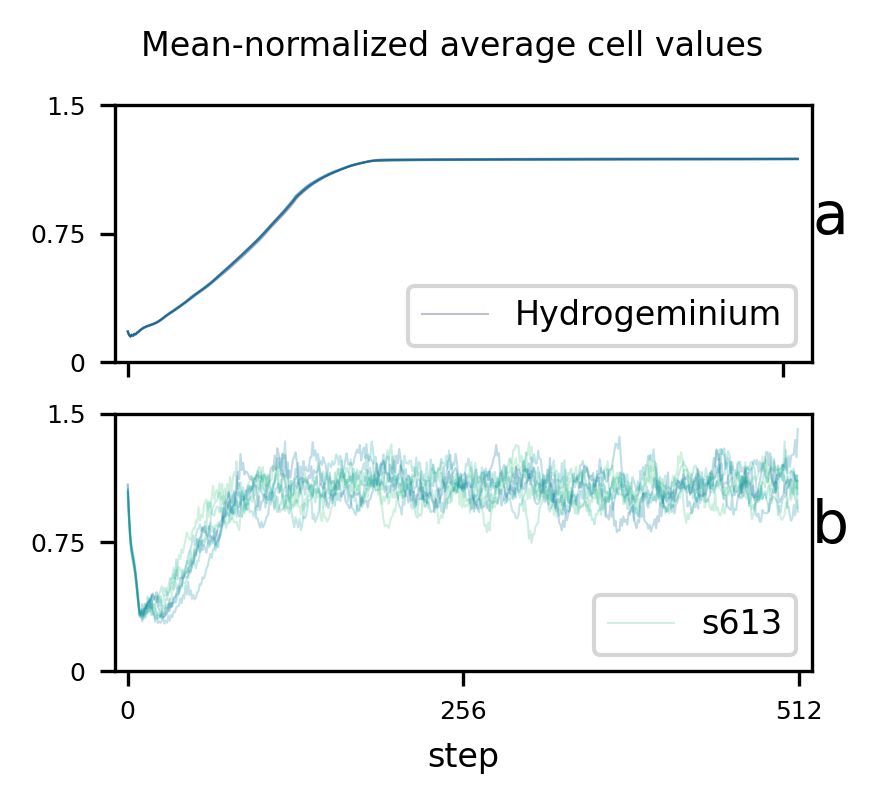}
  \caption{Mean-normalized mean cell value over time. b) After 256 time steps, the normalized average cell value for s613 varies from its overall mean value by a standard deviation of 0.0947, while Hydrogeminium (a) has a standard deviation of 0.00204. As discussed in \citep{wolfram1984}, class III CA are characterized by chaos but tend to an equilibrium cell density, which we don't see in the case of s613.}
  \label{fig:average}
\end{center}                                                            
\end{figure}

\cite{wolfram1984} suggested a subjective classification scheme for complexity in 1D CA, later applied to 2D CA in \citep{packard1984}, and often referenced to as a general rubric for CA. Under the Wolfram classes, briefly, Class I CA cells quickly reach a uniform state (usually all 0), Class II CA typically produce an equilibrium grid of static and/or oscillating patterns, Class III CA produce chaotic dynamics with aperiodic patterns, and Class IV CA exhibit long-lived, complex, and localized patterns. These categories remain subjective, but the sedentary patterns produced by Hydrogeminium might fall under Class II, while the open-ended dynamics of s613 could be categorized as Class III or class IV. Class III CA are suggested in \cite{wolfram1984} to tend toward a consistent density value, but in Figure \ref{fig:average} we see substantial variation in average cell value (normalized to the mean) in s613 over 8 runs, much more than for Hydrogeminium. 

\cite{eppstein2010} introduced a set of heuristic requirements for predicting complexity and universality in CA: mortality and fertility. CA are {\it mortal} if a pattern can be found that disappears after several time steps, and {\it fertile} if a pattern exists that displays the capacity for infinite growth (estimated by finding patterns that escape an initial bounding box). Unlike Wolfram's CA classes and their application to 2D CA \citep{wolfram1984, packard1984} these metrics are not subjective and capture more of the Life-like CA that support universal computation. 

Although fertility and mortality were not developed with continuous CA in mind, Hydrogeminium and s613 meet both criteria. 
Random uniform conditions typically escape an initial bounding box, as demonstrated in Figure \ref{fig:gem_s613_compare} (bounding box, in gray, is twice the width and height of the initialization area). A vanishing pattern for either CA is trivial: a single active pixel with a value smaller than any update function zero-crossing must vanish. Starting from a random uniform initial state (range 0.0 to 1.0), Hydrogeminium typically settles into a mostly static state resembling a Turing pattern \citep{turing1952}. CA s613 remains dynamic, continuously remodeling local structures.

An important motivation for studying artificial life is to explore how to recognize living systems vastly different than found in Earth ecosystems. 
One approach is to take an abstract, information theoretic approach to defining the activities of life. In broad strokes, we can look for systems that generate local departure from thermal equilibrium \citep{popescu2011} and generate statistically unlikely structures. Entropy-based measures find common application in analyzing life and computation alike. Indeed many of the key activities of biological life including replication, transcription, and translation have been analyzed as computation with respect to their thermodynamic efficiency \citep{kempes2017}. Cellular automata, important models of both computation and artificial life, have been been subject to several entropy-based metrics including input entropy \citep{wuensche1999}, local entropy \citep{helvik2004}, conditional entropy \citep{pena2021}, and transfer entropy \citep{lizier2008}.

\begin{figure}[ht]
\begin{center}  
  \includegraphics{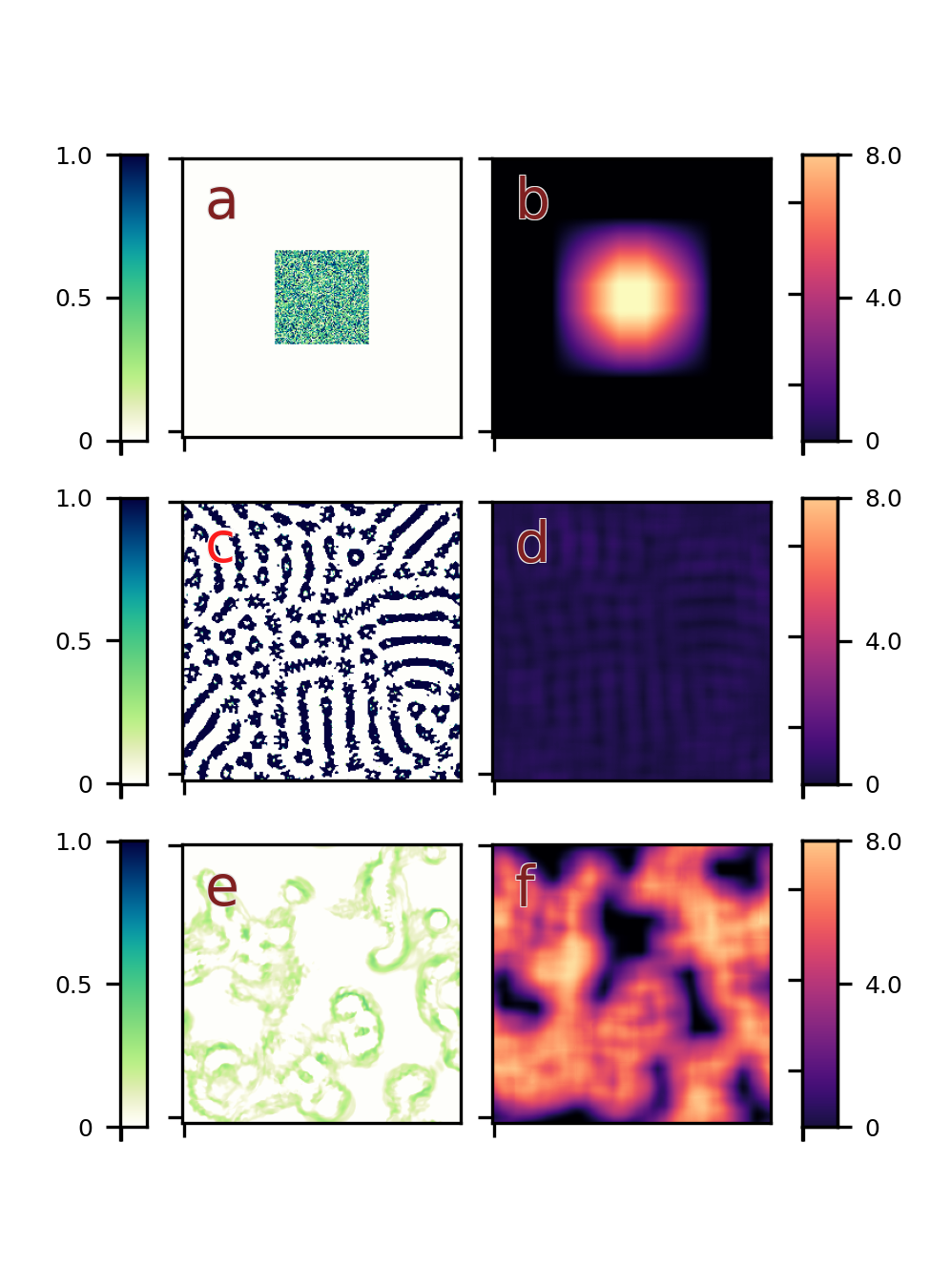}              
  \caption{Spatial entropy maps of step 512 of CA simulation, starting from a bounded random uniform initial grid. a) Random uniform initial grid state. b) Spatial entropy for a. c) Hydrogeminium grid state after 1024 steps and d) spatial entropy of c. e) Grid state of s613 after 1024 steps and f) corresponding spatial entropy map.}
  \label{fig:spatial_entropy}                                                   
\end{center}                                                                
\end{figure}

We consider spatial entropy of CA grids, calculating the image entropy under a sliding window with the same dimensions as the convolution kernel used to compute neighborhood values. Given the neighborhood kernel window size, our spatial entropy measure is conceptually similar to the input entropy in \citep{wuensche1999}, which considers the discrete states of cells defining 1D CA rules of varying input size. The 2D spatial entropy used here considers the cell values (discretized to 8 bits for computational tractability) in the neighborhood kernel window. In the window around each cell in the CA grid we compute entropy as for a 2D 8-bit image.

\begin{equation}
H = -\sum_{s=0}^{S}{ P(s) \log_2(P(s))}
\label{eqn:entropy}
\end{equation}

where $H$ is the entropy of the sub-grid and $P(s)$ is the empirical proportion of cells in the sub-grid in state $s$. 
Computed at each cell location (or pixel), Equation \ref{eqn:entropy} yields a spatial entropy map. In Figure \ref{fig:spatial_entropy} spatial entropy is visualized for a random uniform initial state (confined to a starting box), and after several hundred time steps of change under Hydrogeminium and s613 rules, starting from the same random uniform initialization. After 1024 steps, Hydrogeminium has very little spatial entropy variance while s613 has higher average entropy and much higher variance. Over 8 runs with bounded random uniform initial cell states, Hydrogeminium had final average spatial entropy of 1.18 bits $\pm$ 0.065 (standard deviation), while s613 had final average spatial entropy of 3.95 bits $\pm$ 1.25. 

\begin{figure}[ht]                                                               
\begin{center}                                                            
  \includegraphics{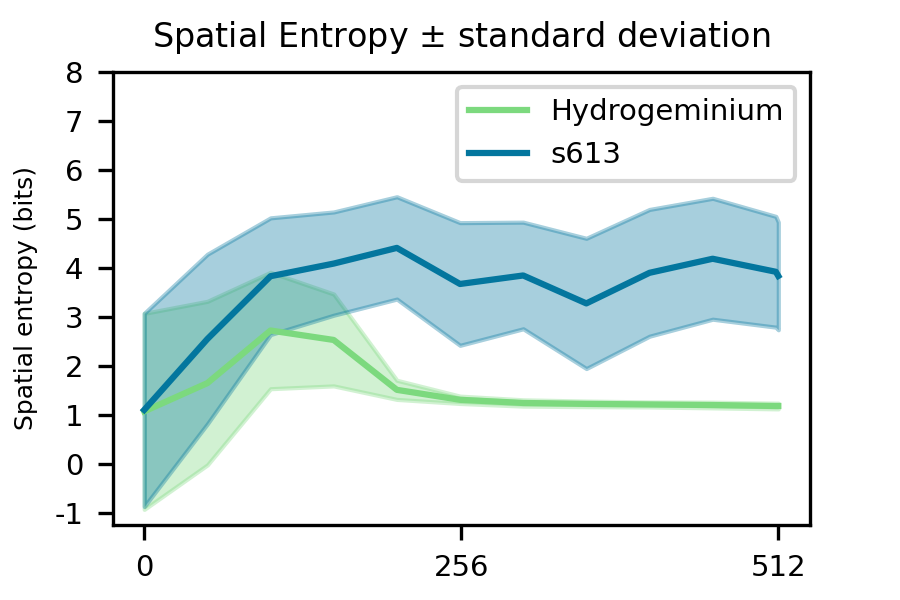}
  \caption{Spatial entropy over time. Values represent 11 snapshots taken every 25 time steps in Hydrogeminium and s613 CA simulations. Starting grid states were identical and sample from a random uniform distribution. Error fill is $\pm$ standard deviation. }
  \label{fig:spatial_entropy_plot}                                                   
\end{center}                                                                
\end{figure}

In addition to variation across the CA grid at a given time step, s613 spatial entropy varies substantially over time, especially as compared to Hydrogeminium in Figure \ref{fig:spatial_entropy_plot}. The upper and lower bounds in Figure \ref{fig:spatial_entropy_plot}, representing standard deviation from the mean, remain consistent and wide over time in s613. Hydrogeminium approaches a static mean value and vanishingly small standard deviation of spatial entropy after about 200 time steps.

\begin{figure}[ht]                                                               
\begin{center}                                                  
  \includegraphics{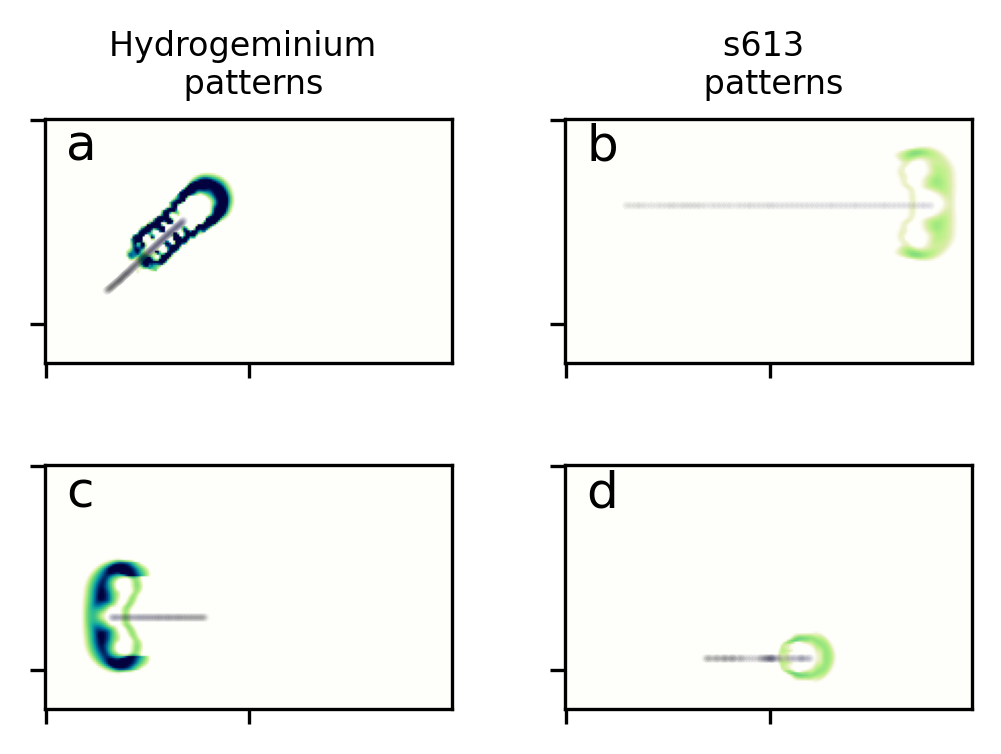}  
  \caption{Self-organizing, traveling patterns in Hydrogeminium and s613. a) Large, slow-moving, and eponymous pattern in Hydrogeminum natans. b) Fast-moving glider in s613. This pattern begins to wobble afer some time steps and eventually becomes unstable. It has a similar pattern in c) glider in Hydrogeminium natans. This pattern starts to wobble after a few hundred time steps, and sometime becomes unstable over long time periods. d) "Frog" pattern in s613. This pattern moves in a straight, hopping motion. Trajectories correspond to 64 time steps for all patterns, except a) which shows 128 time steps.}
  \label{fig:gem_s613_patterns}                                 
\end{center}                                                               
\end{figure}

Mortality and fertility heuristics, subjective categorization, and entropy-based metrics are all attempts to measure and/or predict complexity and universality in CA. They do not fully substitute for finding coherent, self-organizing, traveling patterns known as gliders (or, often, particles in 1D CA). They may serve as effective tools for finding gliders, however, especially when information theoretic metrics are used as filters to highlight the presence of gliders/particles \citep{lizier2008, shalizi2006, wuensche1999, helvik2004}. In 2D CA gliders often are readily apparent from observation, as in the discovery of the classic reflex glider in Life \citep{berlekamp2004}. In this work we present a few examples of gliders in Hydrogeminium and s613, found by evolving synthesis patterns encoded as compositional pattern producing networks \citep{stanley2007} in \citep{davis2022}. These patterns were selected for a combination of center-of-mass displacement and consistent average cell value (a proxy for homeostasis).

Traveling glider patterns from each CA are shown along with trajectories in Figure \ref{fig:gem_s613_patterns}, and include a slow-moving ``wrinkled cucumber" in Hydrogeminium and a hopping ``frog" pattern in s613. Both CA support similar broad gliders that initially have a straight traveling behavior, but begin to wobble after several hundred time steps and often eventually become unstable.

Figure \ref{fig:teaser} shows the Figure \ref{fig:gem_s613_patterns}d pattern emerging from s613 dynamics, and corresponding spatial entropy, with the same grid state simulated for an equal number of steps in Hydrogeminium for comparison. This pattern emerges comparatively often from random uniform initial states in s613, frequently created and destroyed in collisions with other active cells. 


\section{Discussion}

\label{sec:discussion}

By extending Conway's Life to continuous values and time steps, Lenia introduced a combinatorial increase in possible configurations. By focusing on Life, Lenia also introduced a significant simplification over Life-like CA, opting for a single growth function instead of mirroring the cell value-dependent birth and survival rules of Life-like CA. We showed that this simplification limits Lenia's ability to implement arbitrary Life-like CA: there is no straightforward implementation of Life-like CA with birth rules that are not a subset of its survival rules\footnote{But note that Lenia can readily implement Life, which in turn can simulate any other Life-like CA. See for example Brice Due's OTCA metapixel \url{https://otcametapixel.blogspot.com/}}.

Aside from the lack of straightforward implementations of many of the 262,144 possible Life-like CA in Lenia, any significant limitation as a model for artificial life as a result of the growth function simplification in Lenia is not readily apparent: work with the framework has generated a large taxonomy of bioreminiscent patterns \citep{chan2019} and Lenia has continued to be the substrate for increasingly automated exploration of bioreminiscent patterns \citep{reinke2020, davis2022} and modifications to the framework itself \citep{chan2020, kawaguchi2021}. Nonetheless, there is a vast number of Life-like CA with non-overlapping and/or non-contiguous B/S rules, including the Morley/Move CA we focused on in this article and the B356/S23 CA found in \cite{pena2021} to score the highest on a complexity metric based on conditional entropy\footnote{Like Morley, B356/S23 has a frequently-occurring puffer pattern, a reflex puffer with period 72}, and the increased rule-space enabled with Glaberish likely contains many interesting CA. 

In comparing related CA in Lenia and Glaberish, we observed that Hydrogeminium tends to eventually produce mostly sedentary Turing patterns, an equilibrium characteristic of class II CA in Wolfram's classification scheme. The Glaberish CA s613 continuously remodels local structures and remains difficult to predict, characteristic of Class III or IV (chaotic or complex) CA under Wolfram's classification \citep{wolfram1984}. The markedly different dynamics of the two related CA make predicting future states in Hydrogeminum significantly easier than in s613. It is not clear how much the framework influences the CA dynamics in comparison to the way each rule set was developed: s613 was evolved to be difficult for convolutional neural networks to predict whether a given pattern would disappear after a number of time steps \citep{davis2022}, whereas Hydrogeminium was developed with a combination of manual and semi-automated evolution \citep{chan2019}. Also, not all Lenia CA share the characteristic end-point of a static Turing pattern\footnote{Examples of more dynamic Lenia CA include {\it Astrium scintillans} and two variants of {\it Pentafolium incarceratus}.}.

We also considered variance in spatial entropy as a life-like characteristic for assessing CA. One hallmark of life seems to be unusual or improbable structure, {\it e.g.} living eukaryotic cells maintain an improbable distinction between the external environment and their internal milieu, as well as distinct environments within organelles that quickly return to equilibrium when life stops. 
In terms of spatial entropy, s613 is more varied over both space and time than Hydrogeminium (Figures \ref{fig:spatial_entropy} and \ref{fig:spatial_entropy_plot}). Spatial entropy considered here is a nearly analogous metric to the input entropy described in \citep{wuensche1999}. Wuensche described a characteristic signature of complex CA in entropy variance, which we also see in s613, across space as well as time. Other information theoretic measures of complexity have been applied to CA, including local entropy \citep{helvik2004}, conditional entropy \citep{pena2021}, transfer entropy \citep{lizier2008}, and local sensitivity and statistical complexity \citep{shalizi2006}. While not considered here, the project repository includes tools for computing conditional entropy on continuous CA.

We also demonstrated that both Hydrogeminium and s613 support gliders. Gliders form a foundational basis of computation in CA, readily performing the essential functions of information transfer by their ability to travel, and information modification in the consequences of their collisions \citep{berlekamp2004, lizier2008, lizier2010}. Gliders are essential components of engineered computing devices in CA, such as Turing machines \citep{rendell2011} or Life metacells capable of simulating arbitrary Life-like CA\footnote{See, for example, Brice Due's OTCA metacell \url{https://otcametapixel.blogspot.com/}}. 

Information theoretic filters have been shown to be effective for identifying gliders and other coherent structures in CA \citep{wuensche1999, shalizi2006, helvik2004, lizier2008}. Our gliders, in contrast, were obtained from synthesis pattern evolution, selected for center-of-mass displacement and mean cell value homeostasis \citep{davis2022}.


\section{Conclusions}

\label{sec:conclusions}

In this paper we discussed a potentially important limitation of the Lenia framework: updates are applied without regard to current cell values, which limits facile implementation of Life-like CA to those with birth rules wholly contained as a subset of the survival rules. Life, with its B3/S23 rule set, can be readily implemented in Lenia, but not Morley or other CA with non-overlapping B/S rules. We demonstrated that by splitting the update function into conditional genesis and persistence functions (analogous to B and S rules),  the ability to implement Life-like CA with distinct B/S rules can be recovered, as shown for the Morley puffer pattern.

In a direct comparison between related CA implemented in the Lenia and Glaberish frameworks, the Glaberish CA is more active and unpredictable. Both CA meet the mortality and fertility criteria predicting universality from \citep{eppstein2010}, but s613 exhibits greater variance in entropy across space and time.

More importantly, both CA support gliders. Gliders are important components of computational ability, and drivers of life-like characteristics, in CA because they transmit information and can perform computation in the results of their collisions. Substantial efforts have been made to develop formal classification or heuristics that are predictive of CA that are computationally universal and compelling, but there seems to be no comprehensive substitute for finding gliders.


As noted in \citep{packard1984}, increasing degrees of freedom in CA can increase the tendency to chaos, and as noted in \citep{chan2020} expanding CA to greater numbers of channels and dimensions makes persistent self-organizing patterns rare, but potentially more interesting. Continuously-valued CA implementations certainly fall under ``additional degrees of freedom'' as compared to their discrete counterparts. 
In addition, Glaberish at least doubles the number of functions used to calculate updates, but if applied to the Lenia's expanded universe extension \citep{chan2020}, where multiple channels may interact in different ways, Glaberish may lead to an exponential increase in rule functions. Determining whether the trade-off of a more complicated implementation is worth the greater expressiveness of the Glaberish formulation, specifically whether the framework is meaningfully more capable for a given task, remains the subject of future work. While continuous dynamics and strong localization of structure that we demonstrated for the Glaberish automaton s613 are promising, our comparisons were limited to one pair of exemplary (and related) CA from the Lenia and Glaberish frameworks and do not claim superiority of one CA framework over the other.

Similar arguments for parsimony could be made against Lenia and other extended CA frameworks with additional degrees of freedom, compared to simple CA like Life. Even Life may seem too complicated if we also have simpler systems available, like the Turing-complete rule 110 elementary CA \citep{cook2004}. 

A strong argument in favor of more complicated CA is the potential for agents to exist fully embodied in simulated environments that follow the same physics as the agents themselves. Continuous CA, complicated as they may be in comparison to their simple antecessors, may fill the gap between real-world robots and simulated agents. In typical simulated learning environments, {\it e.g.} in reinforcement learning research, there is a sharp distinction between the mechanics of control policies and that of the simulated environment. Recent work, described online\footnote{https://developmentalsystems.org/sensorimotor-lenia/} demonstrated a hint of the potential for embodied agents under consistent physics in continuous CA. In a special version of Lenia's expanded universe, the authors used gradient descent methods to optimize glider-supporting CA to improve the gliders' ability to survive interaction with obstacles in an immutable channel. Several CA in that work each give rise to a particular glider pattern that by all appearances skirts obstacles contained in the immutable layer. 


Spatial entropy maps with window size 23 are visualized in Figure \ref{fig:teaser} for a comparison of Hydrogeminium and s613, with s613 exhibiting a glider pattern emerging from CA dynamics. 
An aim of future work is to discover more complex patterns with compartmentalization of internal structures and a clear distinction in internal/external entropy values (or another information theoretic complexity measure), making for an intriguing analog to biological cells.

Glaberish expands the search space of possible CA in Lenia in the same way as Life-like CA expanded the possible CA rules beyond Life itself, providing a more expansive substrate for developing agency and autopoiesis in continuous CA. 
While aesthetics and human pareidolia play a role in the attractiveness of continuous CA patterns, they have the potential to offer increasingly life-like systems for studying life-like computation.




\subsection{Funding} This work was supported by the National Science Foundation under the Emerging Frontiers in Research and Innovation (EFRI) program (EFMA-1830870).



\footnotesize

\bibliographystyle{apalike}
\bibliography{example} 

\end{document}